\documentclass{PoS}
\usepackage[comma,square,numbers,sort&compress]{natbib}
\usepackage{graphicx}
\renewcommand{\baselinestretch}{1.4}

\newcommand{\sqsNN}{\ensuremath{\sqrt{s_{\mbox{\tiny NN}}}}\,}
\newcommand{\pT}{\ensuremath{p_{\mbox{\tiny T}}}\,}

\newcommand{\GeVc}{\ensuremath{\mbox{GeV}/c}\,}

\newcommand{\ee}{$e^{+}e^{-}$\,}

\title{Performance of Calorimetry in ALICE}

\ShortTitle{Performance of Calorimetry in ALICE}

\author{\speaker{Yuri Kharlov}\thanks{for ALICE collaboration}
  \thanks{also in NRC "Kurchatov Institute", Moscow, Russia and MIPT,
    Dolgoprudny, Russia}\\
  NRC ``Kurchatov institute'' -- IHEP, Protvino, Russia\\
        E-mail: \email{Yuri.Kharlov@cern.ch}}

\abstract{The ALICE experiment at LHC studies the
  strong interaction sector of the Standard Model with pp, pA and AA
  collisions. Within the scope of the physics program, measurements of
  photons, neutral mesons and jets in ALICE are performed by two
  electromagnetic calorimeters. Precise and high-granularity photon
  spectrometer (PHOS) composed of lead-tungstate crystals, along with a
  wide-aperture lead-scintillator sampling calorimeter (EMCal) provide
  complementary measurements of photon observables in a wide kinematic
  range. The calorimeter trigger system allows the experiment to utilize
  efficiently the full delivered luminosity, recording a data sample
  enhanced with high-energy photons and jets. 
  Performance of the ALICE calorimeters from proton-proton to
  heavy-ion collision systems is discussed and illustrated by physics
  results derived from data collected by ALICE with its
  electromagnetic calorimeter system.}

\FullConference{Sixth Annual Conference on Large Hadron Collider Physics (LHCP2018)\\
                4-9 June 2018\\
                Bologna, Italy}

\begin{document}

The electromagnetic calorimeter system of the ALICE experiment at LHC
consists of two detectors: a high-precision photon spectrometer PHOS
\cite{Dellacasa:1999kd} and a wide-aperture electromagnetic
calorimeter EMCal~\cite{Abeysekara:2010ze}. The choice of
the parameters of the calorimeters is driven by the physics objectives
which the detectors were designed for. One of the primary goals of the
PHOS is studying thermal properties of the hot strongly interacting
matter created in heavy-ion collisions, by measuring direct photon
radiation at low transverse momenta, \pT. This task relies on photon
and neutral meson detection in a \pT range from hundreds of MeV up to
a hundred GeV. The EMCal was designed for exploring parton energy loss
in the QCD matter via measuring jet spectra, as well as prompt photons
and electrons at high \pT up to 250~\GeVc. The EMCal detector was
installed in the ALICE in two stages: the first part, called EMCal,
was installed in ALICE for the LHC Run1, while an extension of this
calorimeter, referred to as DCal, was install for the LHC Run2.  In
the following both parts are referred to as EMCal, and mention DCal
only when the new modules are in question.  Basic parameters of the
EMCal and PHOS are summarized in Table~\ref{tab:caloparam}.
\newcommand{\defaultbaseline}{\baselinestretch}
\renewcommand{\baselinestretch}{1.0}
\begin{table}[ht]
  \centering
  \begin{tabular}{|p{0.25\hsize}|p{0.34\hsize}|p{0.34\hsize}|} \hline
    & EMCal & PHOS \\ \hline
    Active element & 77 layers (1.44~mm Pb, 1.6~mm
    scintillator) & Crystals PbWO$_4$ \\
    Moli\`ere radius & 3.2~cm & 2.0~cm \\
    Photodetector & APD $5\times5~\mbox{mm}^2$  & APD $5\times5~\mbox{mm}^2$ \\
    Depth         & $20\,X_0$ & $20\,X_0$ \\
    Acceptance    &
         EMCal: $|y|<0.7,~\Delta\varphi=107^\circ$,
         DCal: $0.22<|y|<0.7,~\Delta\varphi=67^\circ$ &
         $|y|<0.13,~\Delta\varphi=70^\circ$ \\
    Granularity &
         Cell $6\times6~\mbox{cm}^2$, &
         Cell $2.2\times2.2~\mbox{cm}^2$, \\
         & $\Delta\varphi\times\Delta\eta=0.0143\times 0.0143$ &
          $\Delta\varphi\times\Delta\eta=0.0048\times 0.0048$ \\
    Modularity &
         EMCAL: 10+2(1/3) modules, DCAL: 6(2/3)+2(1/3) modules, 17664 cells &
         3+1/2 modules, 12544 cells \\ 
    Dynamic range & $0-250$ GeV & $0-100$ GeV \\
    Energy resolution $\sigma_E/E$ &
         $4.8\%/E \oplus 11.3\%/\sqrt{E} \oplus 1.7\%$ &
         $1.8\%/E \oplus 3.3\%/\sqrt{E} \oplus 1.1\%$ \\
    Distance from interaction point & 428 cm, $0.7-0.9\,X_0$ & 460 cm, $0.2\,X_0$ \\
    \hline
  \end{tabular}
  \caption{Basic parameters of the ALICE calorimeters EMCal and PHOS.}
  \label{tab:caloparam}
\end{table}

\begin{figure}[ht]
  \parbox{0.48\hsize}{\includegraphics[width=\hsize]{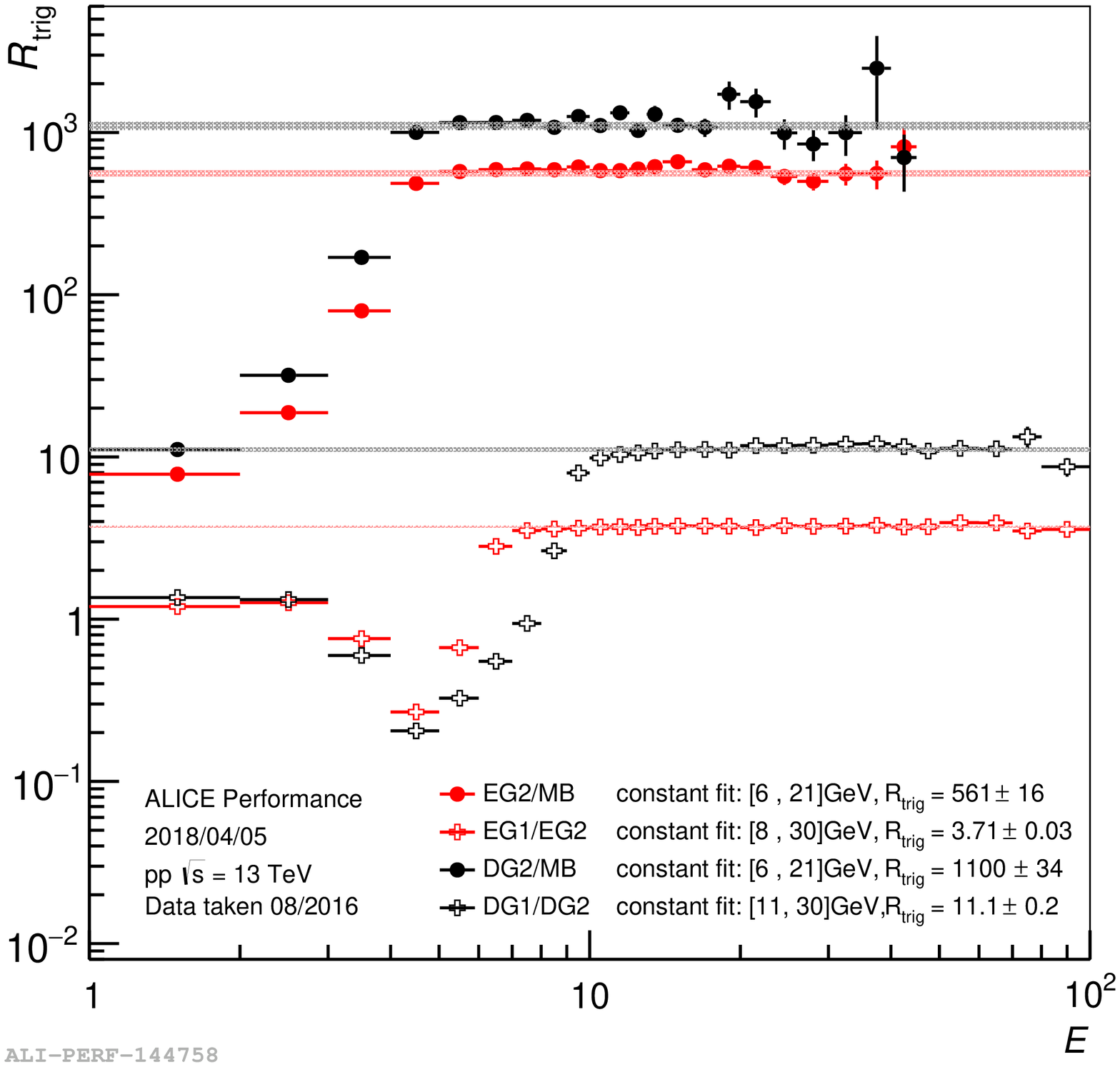}}
  \hfil
  \parbox{0.50\hsize}{\includegraphics[width=\hsize]{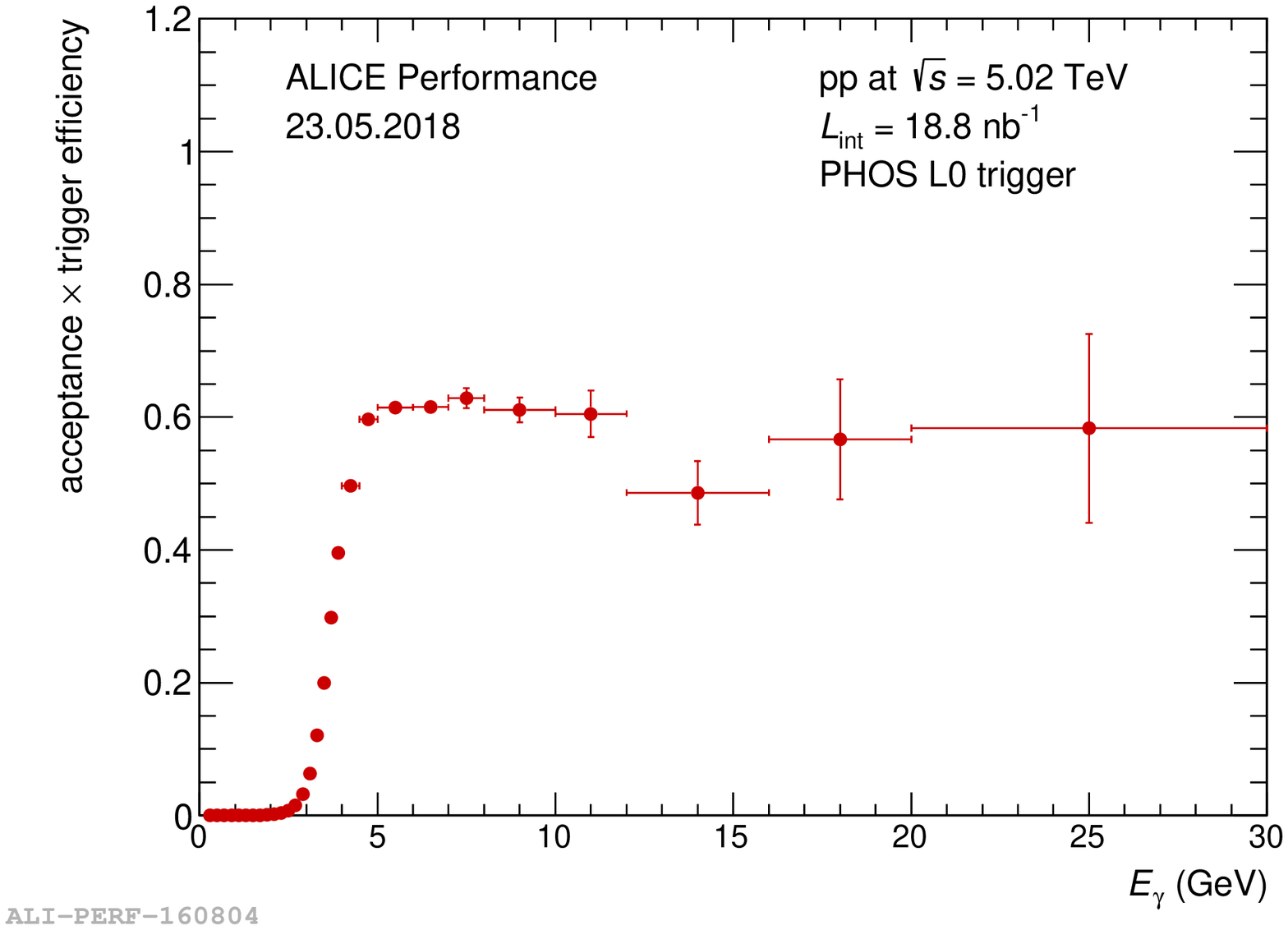}}
  \caption{Turn-on curves for EMCal L1 triggers in pp collisions at
    $\sqrt{s}=13$~TeV and for PHOS L0 trigger in pp collisions at
    $\sqrt{s}=5.02$~TeV. 
  }
  \label{fig:TriggerTurnOn}
\end{figure}

The ALICE calorimeters record collision data together with other
detectors by triggering on minimum-bias beam-beam interactions, but
EMCal and PHOS also can serve as self-triggered detectors by recording
events characterized by high-energy objects detected in the
calorimeters: photons, electrons and jets. These triggering
capabilities on rare events with small cross sections allows the
experiment to inspect the full luminosity delivered by the LHC. The
EMCal and PHOS generate the level-0 (L0) trigger which is fired with a
latency of $1.4~\mu$s after beam interaction by selecting events with
clusters of high energy above a configurable threshold, which is
typically set to 2.5~GeV in EMCal and 4~GeV in PHOS. The calorimeter
trigger system can also generate the level-1 (L1) triggers with a
latency $7~\mu$s by inspecting events preselected by L0 and searching
for events with high-energy photons and jets with more elaborated
algorithms. Several L1 triggers are generated by the
calorimeters. EMCal emits two L1 triggers on photon-like clusters with
two energy thresholds which are equal to 4 and 9~GeV in pp collisions,
and two triggers on jets with thresholds 16 and 20~GeV in pp
collisions. PHOS generates only photon L1 triggers with three
thresholds. Performance of the calorimeter triggers can be represented
by the ratios $R_{\rm trig}$ of the photon spectra measured with the
corresponding triggers to those from the minimum bias trigger or from
the calorimeter trigger with a lower threshold, as shown in the left
plot of Fig.\,\ref{fig:TriggerTurnOn} for the EMCal trigger. Similar
characteristics is the trigger efficiency, the probability of a
high-energy photon to generate the trigger, as illustrated on the
right plot of Fig.\,\ref{fig:TriggerTurnOn} for PHOS.  In EMCal, the
event rates of L1 triggers with low threshold are too high and are
therefore downscaled to accommodate the readout rate accepted by the
ALICE data acquisition. The EMCAL L1 triggers with high threshold and
the PHOS L0 trigger have rather low event rates and can be recorded
without downscaling.

The dominant background in the measurement of direct photons is
determined by photons from light neutral decays, mainly $\pi^0$ and
$\eta$, therefore one of the primary tasks for the ALICE calorimeters
is a precise measurement of neutral meson spectra in a wide \pT
range. Neutral meson spectra measured in pp, pA and AA collisions are
important themselves as a sensitive probe to study hard parton energy
loss in QCD medium via measurement of the nuclear modification factor
$R_{AA}$. The advantage of $\pi^0$ and $\eta$ spectra measurement in the
ALICE calorimeters with respect to charged hadrons is provided by a
much wider energy range accessible for neutral meson identification.

\begin{figure}[ht]
  \parbox{0.48\hsize}{\includegraphics[width=\hsize]{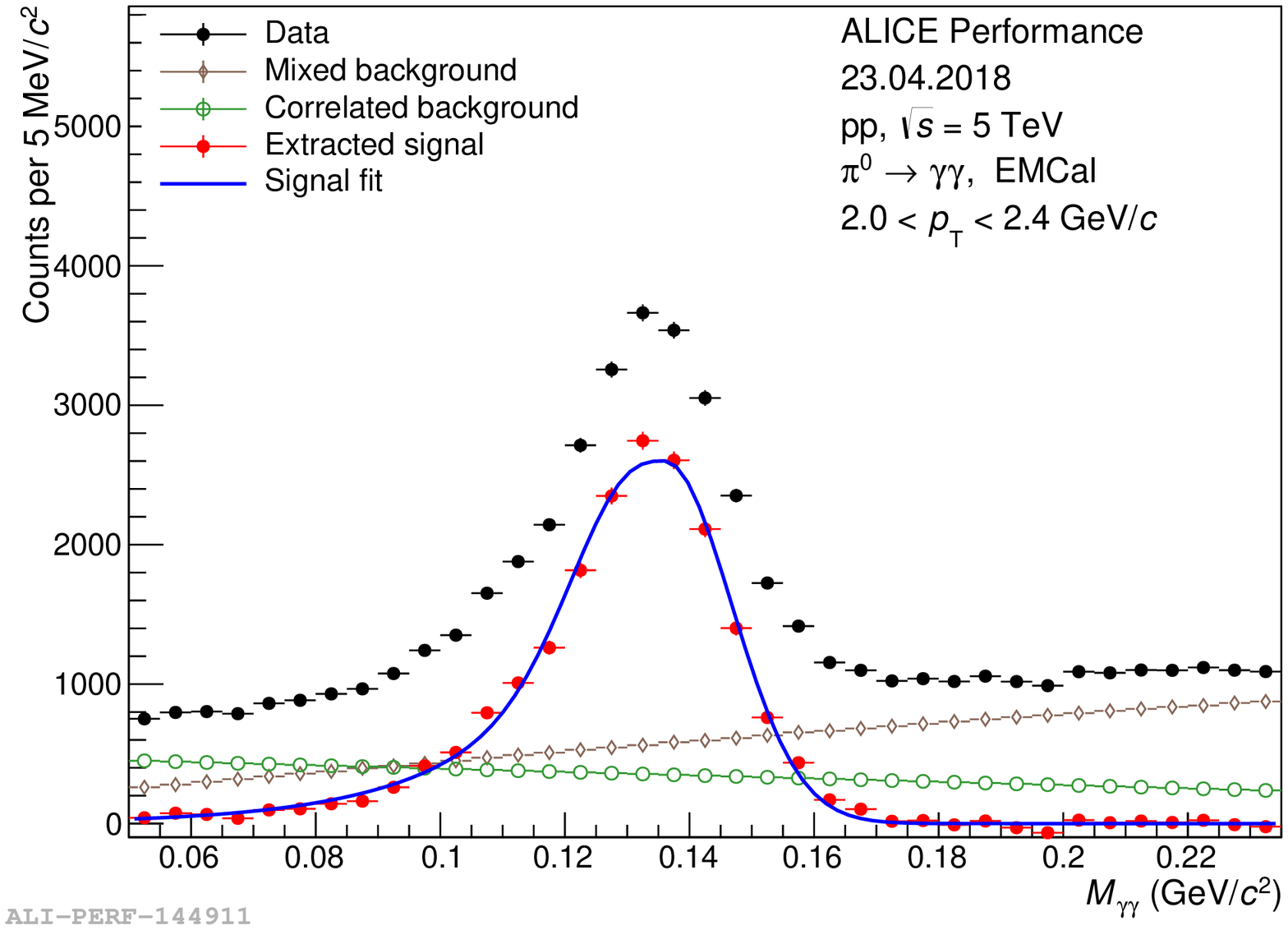}}
  \hfill
  \parbox{0.48\hsize}{\includegraphics[width=\hsize]{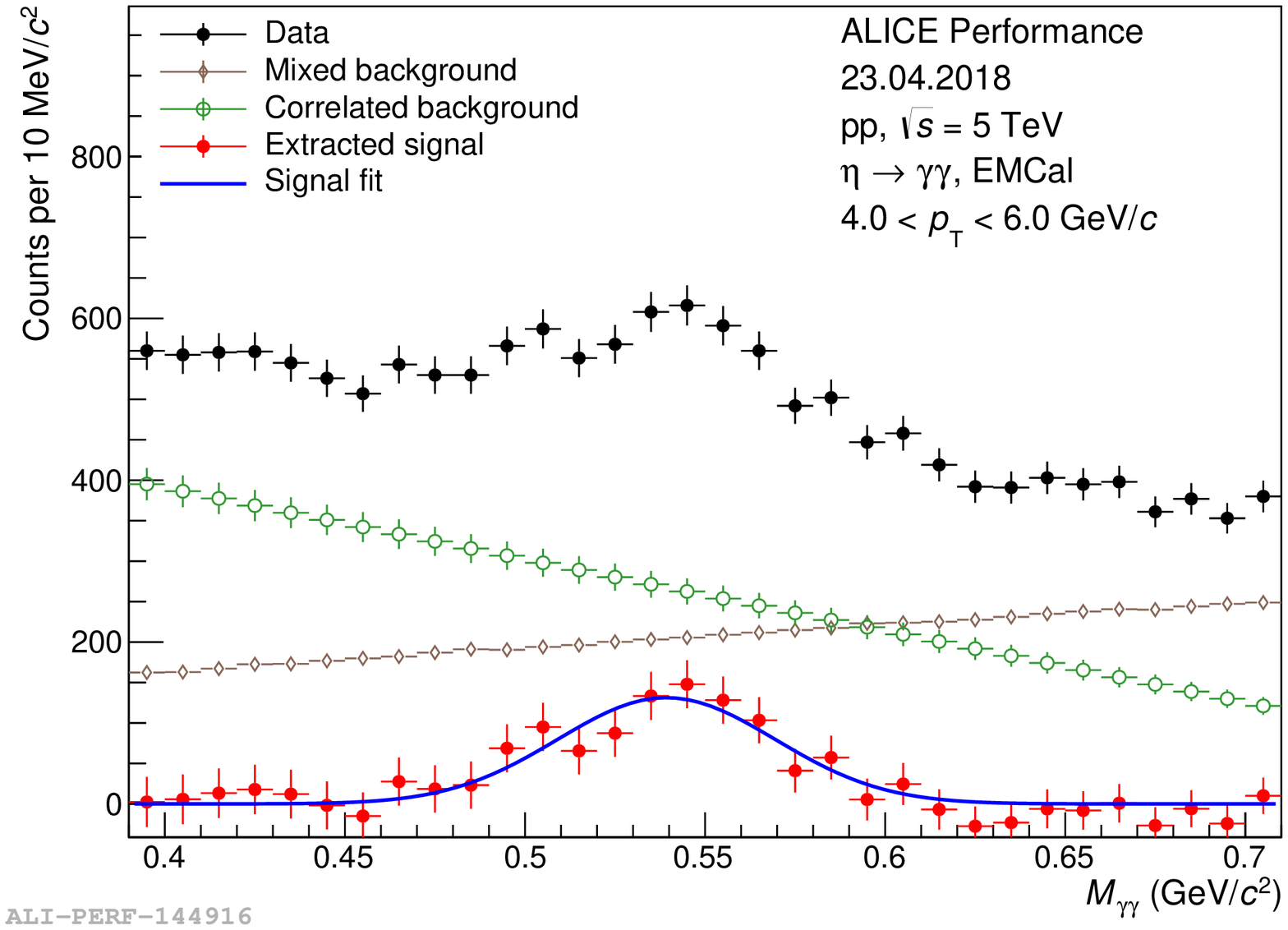}}
  \parbox{0.45\hsize}{\includegraphics[width=\hsize]{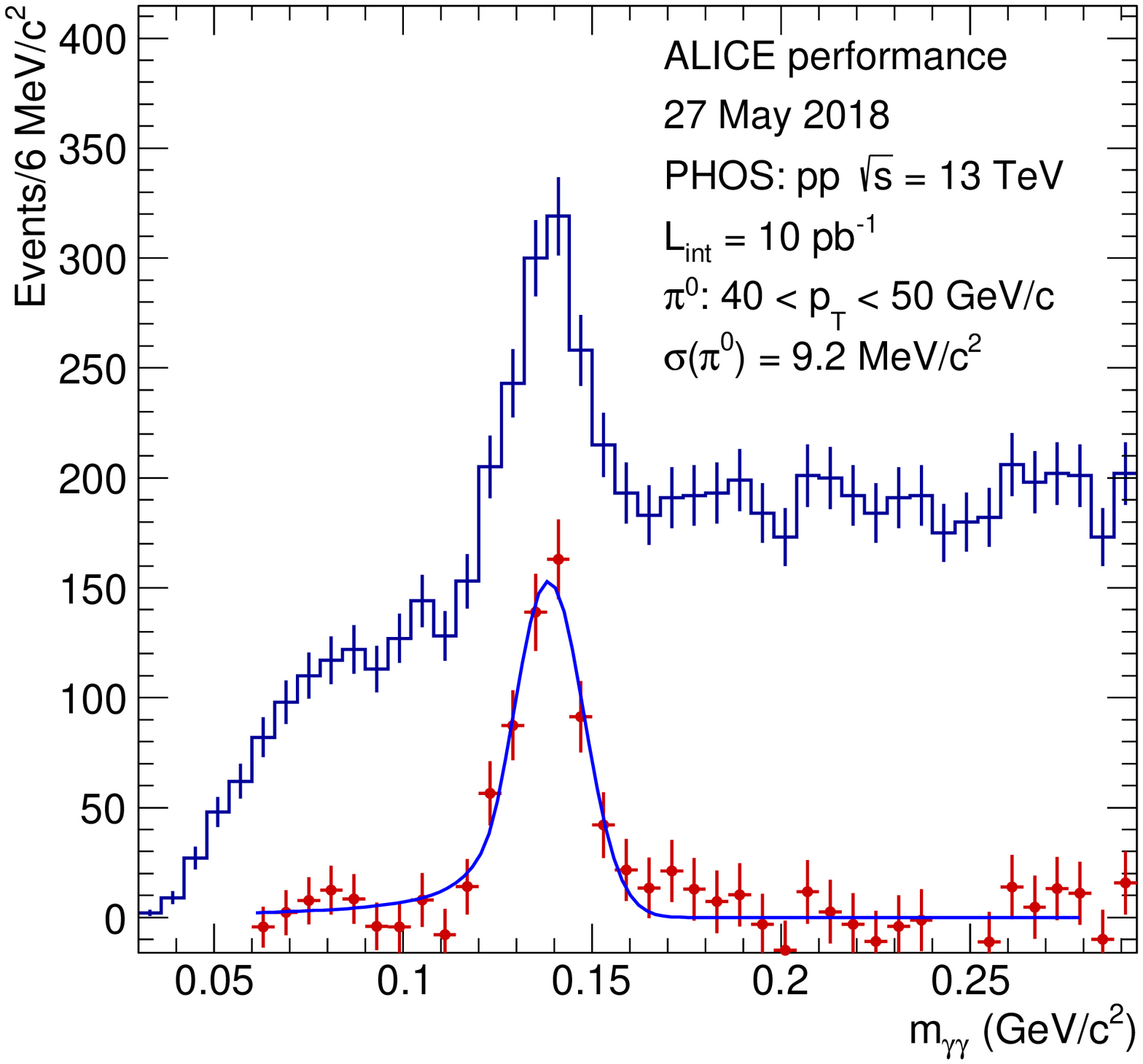}}
  \hfill
  \parbox{0.45\hsize}{\includegraphics[width=\hsize]{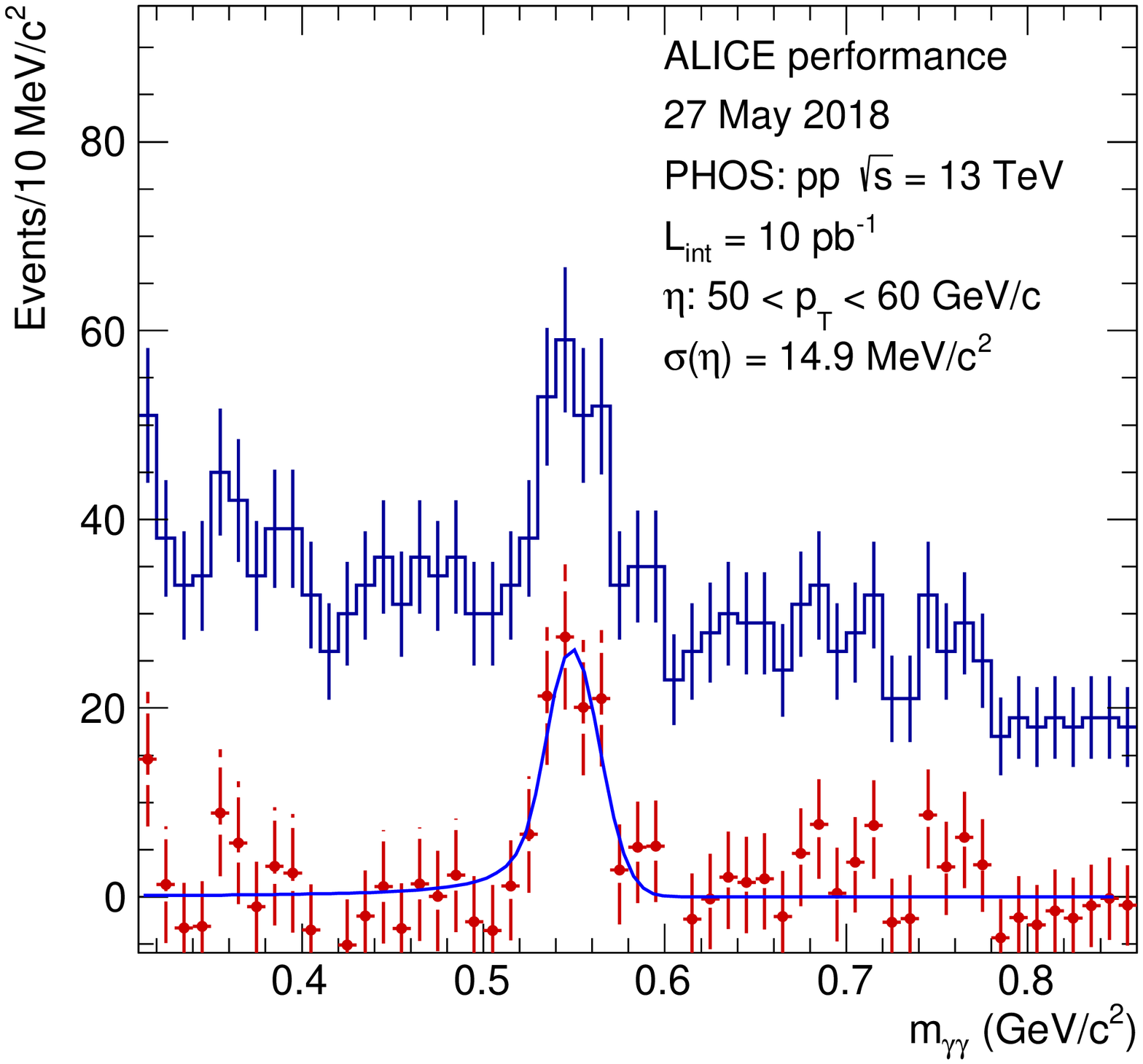}}
  \hfil
  \caption{Two-photon invariant mass spectra in pp collisions at
    $\sqrt{s}=5.02$~TeV in EMCal (upper row) and at $\sqrt{s}=13$~TeV
    in PHOS (bottom row) in the $\pi^0$ and $\eta$ meson peak regions.}
  \label{fig:InvMassSpec_pp}
\end{figure}

The main method of neutral meson reconstruction in EMCal and PHOS is
to build the invariant mass spectra of photon pairs with combinatorial
background subtraction. Fig.\,\ref{fig:InvMassSpec_pp} presents
examples of two-cluster invariant mass spectra in various \pT bins in
pp collisions at 5 and 13~TeV. These spectra show clear peaks
corresponding to $\pi^0$ and $\eta$ mesons above a combinatorial
background. The background has a contribution from correlated cluster
pairs coming from the same source, such as jets or secondary
interaction of photons with the detector material, as illustrated in
Fig.\,\ref{fig:InvMassSpec_pp}. The technique of background
modelling by mixed events is deployed in data analysis to subtract the
background.
The Figure \ref{fig:InvMassSpec_pp} also demonstrates the $\pi^0$ and
$\eta$ peaks in PHOS at the highest \pT accessible in pp collisions at
$\sqrt{s}=13$~TeV in 2016--2017 with integrated luminosity $L_{\rm
  int} = 10~\mbox{pb}^{-1}$. As one can see, statistics of photon
pairs at $40<\pT<50~\GeVc$ is high enough for the $\pi^0$, although
$\pT=50~\GeVc$ is the upper limit of $\pi^0$ reconstruction using
the invariant mass method because the photons from $\pi^0$ decays at
higher \pT cannot be separated due to cluster overlap in PHOS. The
upper limit of $\eta$ meson reconstruction is driven by statistics
only due to a larger opening angle of photons from $\eta$ decay. The
ALICE collaboration has performed measurements of $\pi^0$ and $\eta$
yields at mid-rapidity in pp
\cite{Abelev:2012cn,Abelev:2014ypa,Acharya:2017tlv}, p--Pb
\cite{Acharya:2018hzf} and Pb--Pb collisions \cite{Abelev:2014ypa} at
LHC energies.

\begin{figure}[ht]
  \hfil
  \parbox{0.40\hsize}{\includegraphics[width=\hsize]{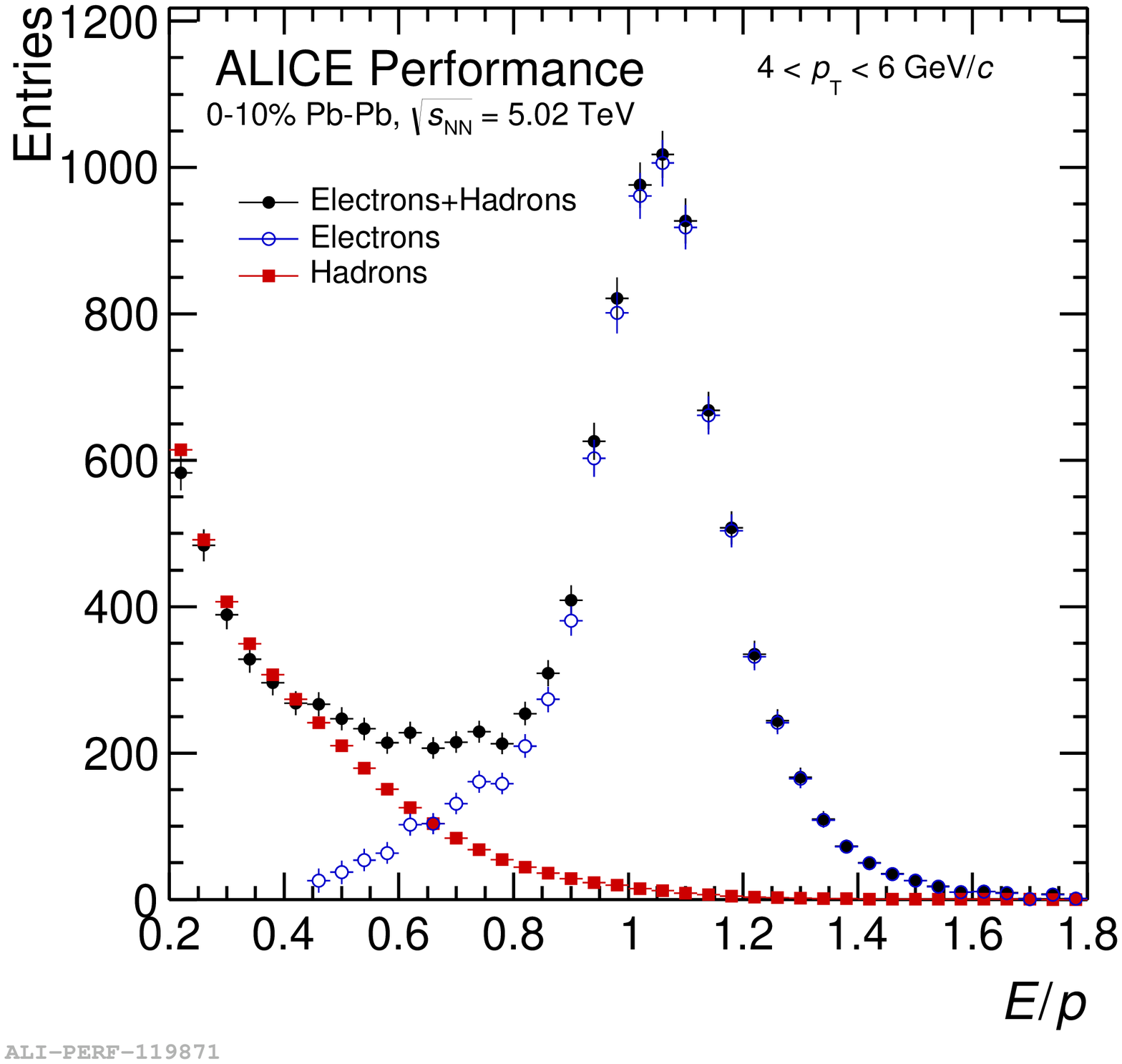}}
  \hfil
  \parbox{0.40\hsize}{\includegraphics[width=\hsize]{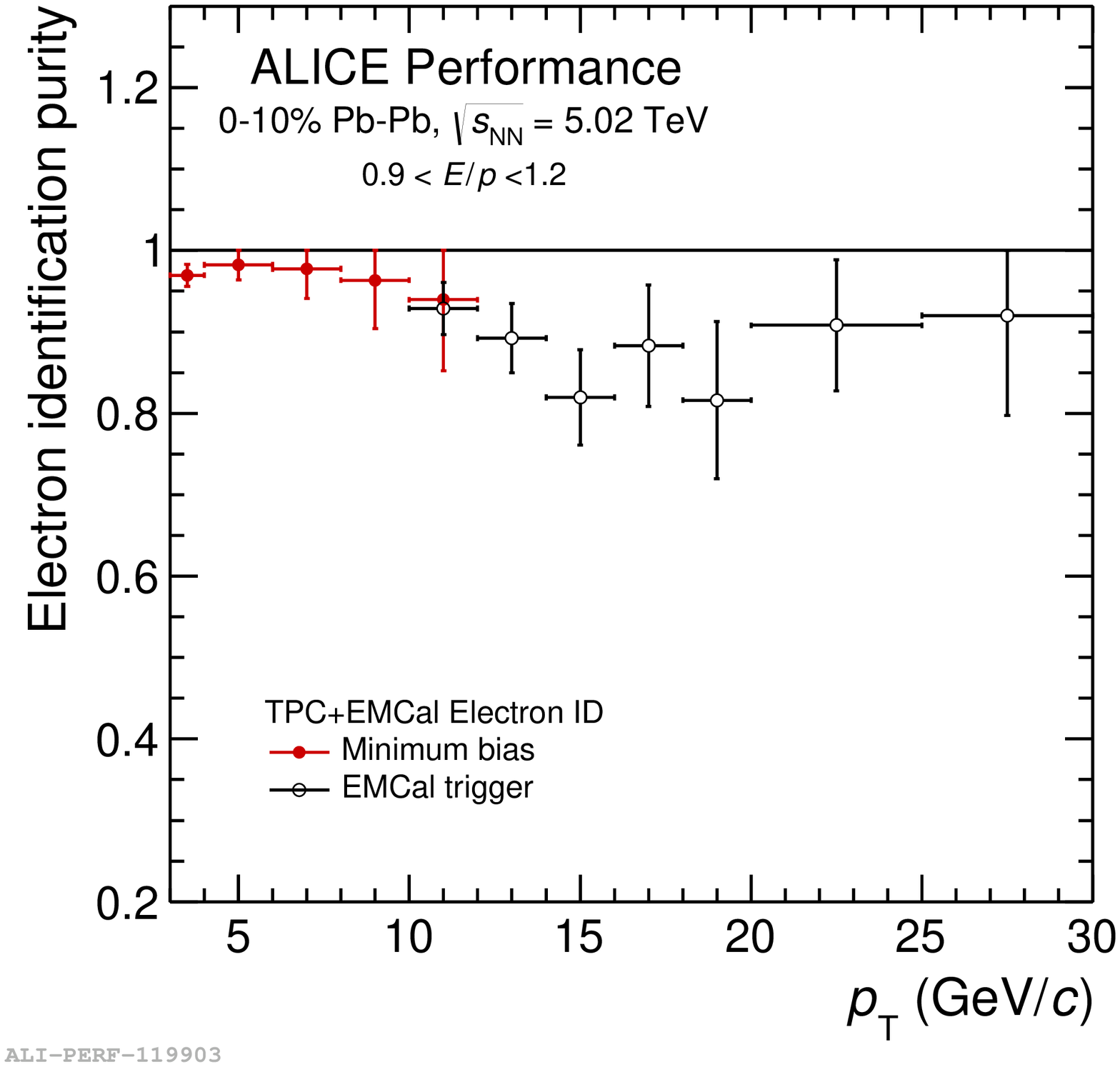}}
  \hfil
  \caption{Distribution of electron energy-over-momentum $E/p$ ratio (left) and electron
    identification purity (right) in central ($0-10\%$) Pb--Pb
    collisions at $\sqsNN=5.02$~TeV in EMCal.}
  \label{fig:EoverP_PbPb}
\end{figure}

Data from the calorimeters, being analyzed together with the ALICE
central tracker, allows one to reconstruct electrons matching the
calorimeter clusters with tracks propagated to the calorimeter
surface. Electrons, unlike other charged particles, deposit their
whole energy in electromagnetic calorimeters, and thus the ratio of
the calorimeter energy $E$ to the track momentum $p$ is equal
to~1. The distribution of $E/p$ ratio of EMCal clusters and matched
tracks in Pb--Pb collisions at $\sqsNN=5.02$~TeV is shown on the left
plot of Fig.\,\ref{fig:EoverP_PbPb}. In conjunction with electron
identification in the ALICE Time Projection Chamber (TPC), EMCal allows
to reconstruct and identify electron in the wide \pT range
$\pT<30~\GeVc$ (Fig.\,\ref{fig:EoverP_PbPb}, right plot).

\begin{figure}[ht]
  \centering
  \includegraphics[width=0.9\hsize]{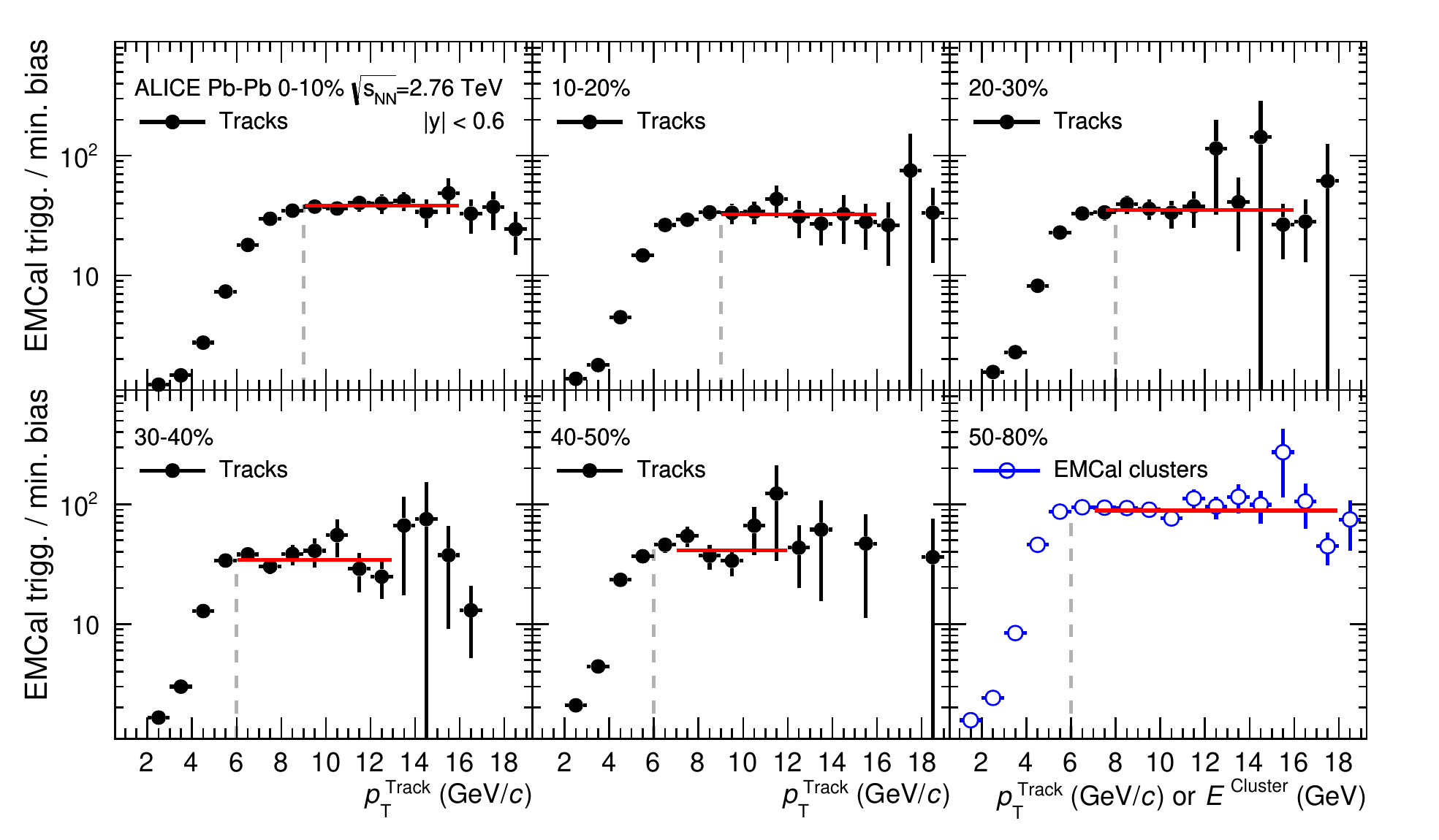}
  \caption{The ratio of inclusive electrons in EMCal triggered events
    to minimum-bias events as a function of associated track \pT in
    Pb--Pb collisions at $\sqsNN=2.76$~TeV in centrality bins from
    $0\%-10\%$ and $50\%-80\%$. The \pT values which separate the spectra
    from the minimum-bias trigger and the EMCal trigger, are
    indicated with black dashed lines.}
  \label{fig:tcurves_electron}
\end{figure}

\begin{figure}[ht]
  \parbox{0.55\hsize}{\includegraphics[width=\hsize]{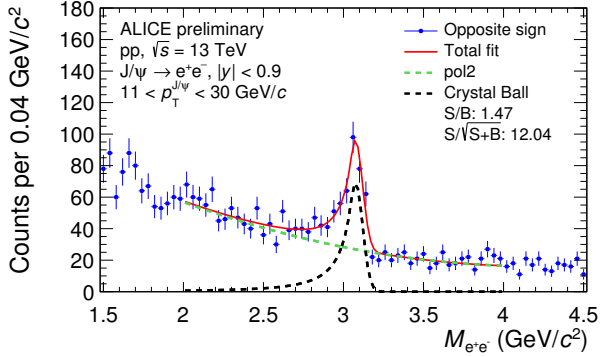}}
  \parbox{0.43\hsize}{
    \caption{Invariant mass spectrum of \ee pairs with $J/\psi$ peak in pp
      collisions at $\sqrt{s}=13$~TeV from EMCal data analysis.}
    \label{fig:InvMassJpsi_pp}
  }
\end{figure}

In measurements of electrons from semileptonic decays of heavy quarks
$c$ and $b$ characterized by small cross sections, the EMCal trigger is powerful
to select events with high-energy electrons. In Pb--Pb collisions at
$\sqsNN=2.76$~TeV the EMCal trigger enhanced electron samples by a
factor of $30-40$ \cite{Adam:2016khe}, as shown in the trigger turn-on
curves of electron \pT spectra in Fig.~\ref{fig:tcurves_electron}.
Electron identification in EMCal is also used for charmonium
reconstruction. Capabilities of $J/\psi$ reconstruction in pp
collisions at $\sqrt{s}=13$~TeV in ALICE are illustrated by \ee
invariant mass spectrum at $11<\pT<30~\GeVc$ in
Fig.\,\ref{fig:InvMassJpsi_pp}, where one electron is reconstructed
and identified in the TPC and at least one electron with $\pT>7~\GeVc$
is reconstructed in EMCal and identified by the energy-to-momentum
ratio $0.8 < E/p < 1.3$.

In the jet measurements, the role of EMCal is to trigger events
with large energy deposited by photons in the EMCal acceptance, as
described in \cite{Adam:2015xea}, and to reconstruct the neutral jet
energy. Charged jet component is reconstructed by the central tracking
system. These jet detection capabilities were used in measurements of
differential cross section of jets in pp collisions at
$\sqrt{s}=2.76$~TeV \cite{Abelev:2013fn}, in jet suppression
measurements in Pb--Pb collisions at
$\sqsNN=2.76$~TeV~\cite{Adam:2015ewa}, in dijet correlations in p--Pb
collisions at $\sqsNN=5.02$~TeV~\cite{Adam:2015xea}.

To summarize, we conclude that the ALICE electromagnetic calorimeters
contribute to the ALICE physics program with photon, neutral meson,
electron, jet measurements in pp, pA and AA collisions. The trigger
system based on the electromagnetic calorimeters allows ALICE to
enhance the collected sample with high-\pT events, and thus utilize
the full luminosity delivered by the LHC. Photon, neutral pion and
electron identification is elaborated in order to ensure high purity
and efficiency.

This work was supported by the RSF grant 17-72-20234.

\bibliographystyle{utphys}
\bibliography{ALICE_calorimetry}

\end{document}